\documentclass[aps,prb,twocolumn,superscriptaddress,showpacs,amsmath, amssymb,amsfonts, floatfix]{revtex4}

\usepackage{graphicx,bm}

\newcommand{\nn}{\nonumber}

\begin{document}
\title{Low-energy Effective Theory for One-dimensional
Lattice Bosons near Integer Filling}

\author{Yu-Wen Lee}
\affiliation{Department of Physics, Tunghai University, Taichung,
Taiwan}

\author{Yu-Li Lee}
\affiliation{Department of Physics, National Changhua University
of Education, Changhua, Taiwan}

\author{Min-Fong Yang}
\affiliation{Department of Physics, Tunghai University, Taichung,
Taiwan}

\date{\today}

\begin{abstract}
A low-energy effective theory for interacting bosons on a
one-dimensional lattice at and near integer fillings is proposed.
It is found that two sets of bosonic phase fields are necessary in
order to explain the complete phase diagram. Using the present
effective theory, the nature of the quantum phase transitions
among various phases can be identified. Moreover, the general
condition for the appearance of the recently proposed
Pfaffian-like state can be realized from our effective action.
\end{abstract}


\maketitle

\section{Introduction}

Much effort has been devoted to understanding the effects of
competing interactions on quasi-one-dimensional systems. Recent
advances in loading ultracold bosonic atoms into an optical
lattice lead to the realization of one-dimensional (1D) lattice
boson systems, and inspire the investigation on many-body quantum
phenomena therein.~\cite{paredes,weiss,stoeferle,fertig} Besides
the nearest-neighbor hopping $t$ and the tunable on-site
interaction $U$, a sizable nearest-neighbor interaction $V$ is now
within experimental reach by using the dipolar interaction among
atoms.~\cite{Griesmaier,molecules} These experiments raise the
interest in creating and detecting exotic quantum phases in 1D
interacting lattice boson systems.

For 1D lattice bosons at integer filling, the phase diagram
obtained by the mean-field calculation includes three different
phases. They are the Mott insulator (MI) for large $U$, a charge
density wave (CDW) for large $V$, and a superfluid (SF) for large
$t$.~\cite{goral,pai} However, by using the Density Matrix
Renormalization Group method, it is found recently that, at
filling of $\bar{n}=1$, there exists a novel insulating phase
lying between the CDW and MI phases in the weak coupling
region.~\cite{Dalla Torre} The phase transition between this new
phase and the conventional insulating phases (i.e., the CDW and MI
phases) is found to be of second order. Similar to the Haldane
phase of quantum spin-one chains,~\cite{haldane83} this new
insulating phase can be identified by a highly nonlocal string
order parameter, and thus called as the Haldane insulator (HI)
phase.~\cite{Dalla Torre} Another investigation employing quantum
Monte Carlo simulations shows that, when $t\ll U\lesssim 2V$,
there exists a supersolid (SS) phase lying between the $\bar{n}=1$
CDW and the $\bar{n}=1/2$ CDW phases.~\cite{Batrouni06} The CDW-SS
transition is found to be of second order with dynamic critical
exponent $z=2$. Recently, due to the possibility of a fault
tolerant quantum computation based on non-abelian
anyons,~\cite{Fault_Tolerant} people begin to search for models
containing non-abelian anyons,~\cite{Models_NAA} as well as for
techniques for their detection and manipulation.~\cite{Detect_NAA}
It is pointed out that bosonic atoms in a 1D lattice with infinite
repulsive three-body on-site interactions around $\bar{n}=1$ have
a ground state very close to a gapless Pfaffian-like
state,~\cite{Paredes06} which may serve as the basis to create
non-abelian anyons in 1D systems.

With an eye on these new developments, it is desirable to have a
unified understanding to the rich phases and the quantum phase
transitions among them through a suitable low-energy effective
theory. A common way of deriving low-energy effective theory for
1D bosonic systems is to use Haldane's density-phase
representation (also known as ``phenomenological
bosonization").~\cite{haldane81,Giamarchi} Such an approach has
been successfully used to address the low-energy physics of SF and
the related Mott transitions at commensurate
fillings.~\cite{Giamarchi} However, this effective theory contains
{\it only one} set of collective variables which describe
long-wavelength phase and density fluctuations, and can show only
one phase transition between two phases.  That is, it fails to
show two phase transitions found in Ref.~\onlinecite{Dalla Torre}.
Moreover, in this simplest one-component hydrodynamic effective
action, the CDW and the SF order parameters compete (or, are dual
to) each other. Therefore, from such an effective theory, it is
hard to understand the origin of the SS phase reported in
Ref.~\onlinecite{Batrouni06}, where the CDW and the SF orders
coexist. In the present work, a low-energy effective theory with
{\it two} sets of bosonic phase fields is proposed. Within the
present framework, not only the nature of the phase transitions
and the corresponding critical modes can be explicitly identified,
the phase diagrams reported in Refs.~\onlinecite{Dalla Torre} and
\onlinecite{Batrouni06} can also be easily understood. Moreover,
we find that the SS phase in 1D has at most an {\it algebraic}
long-range CDW order and has an algebraic {\it boson-paired
superfluid} order instead of the ordinary one. Furthermore, the
general condition for the appearance of the gapless Pfaffian-like
state can be realized within the present approach. This
observation will be of help for searching gapless Pfaffian-like
state in more realistic systems. We believe that the effective
theory shown below provides a general description of the
low-energy physics for 1D interacting lattice bosons near integer
filling, and it serves as a starting point for further
investigations on the static and dynamical properties of these
systems.

\section{The extended boson Hubbard model and its effective theory}

The lattice bosons with a nearest-neighbor hopping $t$, an on-site
interaction $U$, and a nearest-neighbor interaction $V$ can be
described by the extended Bose-Hubbard model (EBHM)
\begin{equation}\label{eqn:bhh}
H = -t \sum_{\langle i,j \rangle} ( b_{i}^{\dag}b_{j} + h. c.) +
\frac{U}{2} \sum_{i} \delta n_{i}^{2} + V \sum_{\langle i,j
\rangle} \delta n_{i}\,\delta n_{j} \, ,
\end{equation}
where $b_{i}^{\dag}$ is the bosonic creation operator at site $i$,
$n_{i}=b_{i}^{\dag}b_{i}$ is the number operator, and $\delta
n_{i} \equiv n_{i}-\bar{n}$ measures deviations of the particle
number from a mean filling $\bar{n}$. Since our main concern is in
discussing the phase transition among various insulating phases,
where the local particle number fluctuations should not be strong,
near an integer filling $n_0$ (i.e., $\bar{n}\simeq n_0$), one can
truncate the bosonic Fock space to only three local states with
particle numbers $n_{0}$ and $n_{0}\pm 1$.~\cite{Dalla
Torre,Paredes06,Altman02,Huber,comment1} Now the lattice boson
model in Eq.~(\ref{eqn:bhh}) can be mapped onto a spin-one chain
in the truncated Hilbert space,
\begin{align}
H_{\rm spin}=& -\frac{n_{0}t}{2} \sum_{\langle i,j \rangle}
(S_{i}^{+}S_{j}^{-} + h. c.) \nn\\
& + V \sum_{\langle i,j \rangle} S_{i}^{z}S_{j}^{z}
+\frac{U}{2}\sum_{i}\left(S_{i}^{z}\right)^{2} \nn\\
&-\frac{n_{0}t\xi}{2} \sum_{\langle i,j \rangle} \bigr[
  S_{i}^{z}S_{i}^{+}S_{j}^{-}+S_{i}^{-}S_{i}^{z}S_{j}^{+} +
  S_{i}^{-}S_{j}^{z}S_{j}^{+}\nn\\&+S_{i}^{+}S_{j}^{-}S_{j}^{z}
 + \xi (S_{i}^{z}S_{i}^{+}S_{j}^{-}S_{j}^{z}+
       S_{i}^{-}S_{i}^{z}S_{j}^{z}S_{j}^{+})
\bigl]  \label{eqn:spinhamiltonian}
\end{align}
due to the correspondence
\begin{equation}\label{eqn:relation}
b_{i}^{\dag} \leftrightarrow \sqrt{\frac{n_0}{2}} \left( 1 + \xi
S_{i}^{z} \right) S_{i}^{+} \; , \qquad \delta n_{i}
\leftrightarrow S_{i}^{z} \, .
\end{equation}
$S_{i}^{\alpha}$ are the $S=1$ spin operators, and the parameter
$\xi\equiv \sqrt{(n_0+1)/n_0}-1$ is a measure of the
`particle-hole asymmetry'. For simplicity, $\xi$ is set to zero,
which is valid at least when $n_0\gg 1$. From the viewpoint of the
bosonization theory, the main physics is indeed not modified by
the effect due to the $\xi$-dependent terms.~\cite{asymmetry} By
further using the composite spin representation for the $S=1$ spin
operators,~\cite{Schulz86,Timonen86} where they are rewritten as a
sum of two commuting spin-1/2 species, the problem of spin-one
chain becomes that with two coupled spin-1/2 chains. After
applying the Jordan-Wigner transformation and bosonization
procedure for these spin-1/2 operators, one finally arrives at the
following bosonized effective
Hamiltonian~\cite{Schulz86,Timonen86,bosonization}
\begin{align}
H_{\rm eff}  & =  \frac{u_s}{2} \int dx \left[ K_s
(\partial_x\Theta_s)^2  +
         \frac{1}{K_s} (\partial_x \Phi_s)^2 \right]
        \nn\\& +  g_1 \int dx
         \cos \left( \sqrt{8\pi} \Phi_s \right)   \nonumber \\
    &  +  \frac{u_a}{2} \int dx  \left[ K_a (\partial_x\Theta_a)^2  +
         \frac{1}{K_a} (\partial_x \Phi_a)^2 \right]
          \nn\\& + \int dx
    \left[ g_2 \cos  \left( \sqrt{8\pi} \Phi_a \right)
     +  g_3 \cos \left( \sqrt{2\pi} \Theta_a \right) \right] \, ,
     \label{Bosonform}
    \end{align}
where $\Phi_{s}$, $\Theta_{s}$ and $\Phi_{a}$, $\Theta_{a}$ are
the symmetric (called as the charge mode) and the antisymmetric
(called as the neutral mode) combinations of the bosonized fields
for the two spin-1/2 operators, respectively. The bosonized forms
of the original lattice boson operators are
\begin{align}
\frac{b_{j}^{\dag}}{\sqrt{a}}&\sim
\sqrt{\frac{n_0}{2}}\sqrt{\frac{2}{\pi a}}
e^{i\sqrt{\frac{\pi}{2}}\Theta_s}\cos(\sqrt{\frac{\pi}{2}}\Theta_a)
+ \cdots 
\; ,\nn\\
\frac{\delta n_j}{a} &\sim \sqrt{\frac{2}{\pi}}
\partial_x\Phi_s - (-1)^{x/a}\frac{2}{\pi a}
\sin\sqrt{2\pi}\Phi_s\cos\sqrt{2\pi}\Phi_a
\label{boson-operator}
\end{align}
with $x=ja$, where $a$ is the lattice constant and plays the role
of a short-distance cutoff. In the derivation of
Eq.~(\ref{Bosonform}), some irrelevant (or less relevant) terms
have been dropped. As a result, the charge and the neutral modes
become decoupled. We note that, while the method of
phenomenological bosonization can give identical form of the
charge part in Eq.~(\ref{Bosonform}),~\cite{Giamarchi} the neutral
mode is missed in that approach. That is the reason why the
conventional theory fails to account for the complete phase
diagram of the interacting lattice bosons near the integer
fillings. As shown in Refs.~\onlinecite{Schulz86} and
\onlinecite{Timonen86}, assuming the couplings in the two coupled
spin-1/2 chains being weak, the values of the parameters in
Eq.~(\ref{Bosonform}) can be obtained as follows: $g_1
=-g_2=(2V-U)/2\pi^2 a$, $g_3 =-n_0 t/\pi a$; the Luttinger
parameters $K_{s}=[1+(U+6V)/\pi n_0 t]^{-1/2}$, $K_a=[1-(U-2V)/\pi
n_0 t]^{-1/2}$; and the sound mode velocities $u_{s} = u_0/K_s$,
$u_a=u_0/K_a$ with $u_0\equiv n_0 ta$. Since the present effective
Hamiltonian is reached under several approximations, the values of
the above parameters should not be treated very seriously. Here we
would like to consider Eq.~(\ref{Bosonform}) as a phenomenological
theory. It provides a universal low-energy physics of the 1D
interacting lattice boson models near integer fillings, such that
the nature of the phase transitions and the long-distance
behaviors of the major correlation functions at each phase can be
predicted. As shown below, this effective Hamiltonian does explain
the results reported recently.~\cite{Dalla
Torre,Batrouni06,Paredes06} It reinforces our belief in the
validity of the present framework.

\section{The phases and phase transitions of the extended Boson Hubbard model}

\begin{figure}
\includegraphics[width=2.4in]{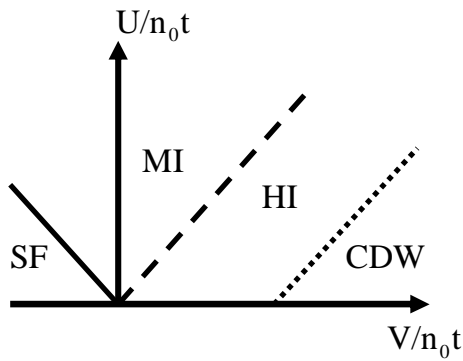}
\caption{The schematic phase diagram at integer filling
$\bar{n}=n_0$ in the weak-coupling regime. The phase boundaries of
Kosterlitz-Thouless, Gaussian, and Ising types are denoted by the
solid, dashed, and dotted lines, respectively.}
\label{fig:fig1}
\end{figure}

Through the discussion on the relevance of each nonlinear term in
Eq.~(\ref{Bosonform}), the phase diagram can be obtained as shown
in Fig.~\ref{fig:fig1}, which looks similar to that reported in
Ref.~\onlinecite{Dalla Torre} in the weak-coupling regime.
According to the work in Ref.~\onlinecite{Schulz86}, better
agreement can be reached if one keeps more local states (say, with
particle numbers $n_{0} $, $n_{0}\pm 1$, and $n_{0}\pm 2$) and
maps the interacting lattice boson model to a spin chain with
higher integer spin.~\cite{note1} However, we must emphasize that
one of the most important feature of the bosonization analysis in
Ref.~\onlinecite{Schulz86} is that the properties of the
corresponding spin model do not dramatically depend on the value
of the spin quantum number $S$, but only on whether $S$ is integer
or half-odd integer. This suggests that the main results obtained
below from our analysis of the Hamiltonian in Eq.
(\ref{Bosonform}) should remain unchanged even when the local
particle number fluctuations become stronger. The phase boundary
separating the SF and the MI phases is determined by $K_s=1$. This
comes from the relevance of the $g_1$ term in the charge sector of
Eq.~(\ref{Bosonform}), and thus gives a type of
Kosterlitz-Thouless transition.~\cite{KT} The MI-HI phase
transition is given by $g_1 =0$. It is a Gaussian transition,
which separates two gapped phases. The HI-CDW transition occurs
when $K_a=1/2$. It is due to the competition of the two relevant
terms (i.e., the $g_2$ and the $g_3$ terms) in the neutral sector
of Eq.~(\ref{Bosonform}), and thus is a Ising-type
transition.~\cite{Ising} By using the bosonized forms of the
original lattice boson operators in Eq.~(\ref{boson-operator}),
the correlation functions at long distance can be calculated. In
the SF phase, because the charge mode is gapless while the neutral
field $\Theta_a$ is pinned, 
the density fluctuation is gapless and the single-particle
correlation function shows an algebraic long range order (LRO)
with an enhanced superfluid correlation: $\langle b_i^\dag
b_j\rangle\sim |i-j|^{-1/4K_s}$. True LRO can develop when
$K_s\rightarrow \infty$. In the CDW phase, both of $\Phi_s$ and
$\Phi_a$ are pinned.
In this case, while the single-particle correlation functions
decay exponentially, the staggered part of the density-density
correlation shows LRO: $\langle \delta n_i \delta n_j\rangle\sim
(-1)^{|i-j|}$. This justifies the identification of the CDW phase.
In the MI and the HI phases, both $\Phi_s$ and
$\Theta_a$ are pinned, 
and both the single-particle and density correlation functions
decay exponentially at long distance. Both in these two gapped
phases, the lattice translation symmetry does not break. The only
difference is that the $\Phi_s$ field is pinned at $0$ and
$\sqrt{\pi/8}$ in the MI and the HI phases, respectively. As
pointed out in Ref.~\onlinecite{Dalla Torre}, one can distinct the
HI phase from the MI one via a nonlocal string order parameter
defined by $\mathcal{O}\sim \left\langle\delta n_i
\exp\left(i\pi\sum_{k=i}^{j} \delta n_k\right)\delta n_j
\right\rangle$. Within the present bosonization approach, such a
string order parameter can be expressed as $\mathcal{O}\sim
\left\langle\sin\left(\sqrt{\frac{\pi}{2}}\Phi_s(x_i)\right)
\sin\left(\sqrt{\frac{\pi}{2}}\Phi_s(x_j)\right)
\right\rangle$,~\cite{Nakamura03} which is completely independent
of the neutral mode. Thus it is indeed nonzero in the HI phase
with $\Phi_s$ being pinned at $\langle\Phi_s\rangle=\sqrt{\pi/8}$,
consistent with the findings in Ref.~\onlinecite{Dalla
Torre}.~\cite{note2} To make the relations between the relevant
couplings and the pinned fields in each phase more explicit, we
summarize the above discussions in table \ref{tab1}.

\begin{table} \tabcolsep=6pt
\begin{tabular}{c|cccc}
\hline\hline
 ~& SF & MI & HI & CDW \\
 \hline
 $K_s$ & $>1$ & $<1$ & $<1$ & $<1$\\
 $K_a$ & $>\frac{1}{2}$ & $>\frac{1}{2}$ & $>\frac{1}{2}$ & $<\frac{1}{2}$\\
 $g_1$ & irrelevant & $<0$ & $>0$ & $>0$ \\
 $g_2$ & irrelevant & irrelevant & irrelevant & $<0$ \\
 $\langle\Phi_s\rangle$ & not pinned & $0$ & $\sqrt{\frac{\pi}{8}}$ &
 $\sqrt{\frac{\pi}{8}}$ \\
 $\langle\Phi_a\rangle$ & not pinned & not pinned & not pinned
 & $0$
 \\
  $\langle\Theta_a\rangle$ & 0 & 0 & $0$
 & not pinned \\
 \hline\hline
\end{tabular}
\caption{The Luttinger parameters and the roles of various
couplings in different phases at integer fillings. The values of
the fields $\Phi_{s, a}$ and $\Theta_a$ indicated in this table
are the values at which these fields were pinned in each phase.
Notice that throughout our discussions, we take the coupling $g_3
<0$ (following Ref.~\onlinecite{Schulz86}), and it is irrelevant
only in the CDW phase above.} \label{tab1}
\end{table}


\begin{figure}
\includegraphics[width=2.4in]{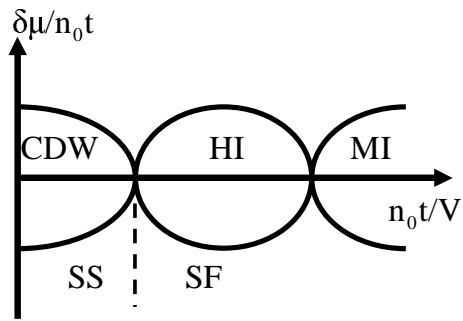}
\caption{The schematic phase diagram near integer filling
$\bar{n}\simeq n_0$. The solid lines denote the
commensurate-incommensurate transition with
$z=2$. 
} \label{fig:fig2}
\end{figure}

Up to now, our discussion has focused on systems exactly at
integer fillings. The deviation from integer filling can be
studied by turning on a nonzero chemical potential $\delta\mu$ and
adding a term $\delta\mu\sum_i \delta n_i$ in the Hamiltonian in
Eq.~(\ref{eqn:bhh}). ($\delta\mu$ is measured from the middle of
the lobe.) In the bosonized representation, this term takes the
form $\delta\mu\sqrt{\frac{2}{\pi}}\partial_x\Phi_s$, which
affects the charge mode only.~\cite{g2_renormalization} The
$\partial_x\Phi_s$ term tends to create a finite number of kinks
in the $\Phi_s$ configuration, and it will compete with the
Sine-Gordon term $\cos\sqrt{8\pi}\Phi_s$ which tends to pin down
$\Phi_s$. When $\delta\mu$ is comparable to the magnitude of the
charge gap of the insulating phases, such a term will trigger a
well-known commensurate-incommensurate
transition.~\cite{Giamarchi} As long as $K_s$ is not too small
such that the breather mass gap is larger than the Sine-Gordon
kink mass gap, the universal features of this transition is well
captured by the Luther-Emergy limit of the charge mode, where the
bosonized effective Hamiltonian can be further mapped to a free
massive Dirac fermion theory with a chemical potential
term.~\cite{Tsvelik} At the tranistion point
$\delta\mu=\delta\mu_c$ such that the charge gap closes, the
gapless charge mode excitation density obeys the scaling relation
$\rho\sim(\delta\mu-\delta\mu_c)^{1/2}$, and the dynamic critical
exponent of this transition becomes $z=2$. After the transition,
the charge mode is described by a Gaussian model with nonzero
stiffness, which accounts for the SF behavior in 1D. Therefore,
starting from either the MI or the HI phase, nonzero $\delta\mu$
can induce a transition to the SF phase. On the other hand, after
the transition from the CDW phase, even though the gapless charge
mode shows the SF behavior, the CDW-like behavior can remain, and
thus the SS phase is obtained. It is because the neutral sector is
unaffected and the $\Phi_a$ field is still pinned in this case
even after the transition. According to this analysis, it is
tempted to interpret the SS phase near integer fillings as the
descendant of the CDW phase at integer fillings. Thus our
effective theory does explain the occurrence of a $z=2$
second-order CDW-SS phase transition found in
Ref.~\onlinecite{Batrouni06}. It is noted that, since the the
charge mode is gapless in the SS phase, a power-law behavior in
the density-density correlation function $(-1)^{|i-j|}
\langle\delta n_i\delta n_j\rangle \propto |i-j|^{-K_s}$ is
expected. That is, the SS phase has at most an {\it algebraic}
long-range CDW order. Because $K_s \ll 1$ in the SS phase, such a
power-law decay may not be examined easily by finite-size
numerical calculations. Another interesting feature of the SS
phase is that, while the superfluid stiffness is nonzero, the
single-particle correlation $\langle b_{i}^\dagger b_{j}\rangle$
decays {\it exponentially} due to pinning of the $\Phi_a$ field
(or the $\Theta_a$ field being disorder). However, the boson-pair
correlation $\langle (b_i^\dagger)^2 b_j^2\rangle$ indeed shows an
algebraic long range order. Thus the system behaves somewhat like
a ``boson-paired superfluid". This issue deserves further
examinations by other approaches. The above discussions are
summarized by the phase diagram shown in Fig.~\ref{fig:fig2}.

\section{The Pfaffian-like states}

Besides the application to the 1D EBHM, a recently proposed
gapless Pfaffian-like state~\cite{Paredes06} as the ground state
of three-body-hard-core bosons in 1D lattice can also be
understood within the present framework. Due to the infinite
three-body on-site repulsion, only three local states with
occupation numbers $n=0$, 1, 2 are necessary. Similar to the
previous discussion, the interacting lattice boson model in
Ref.~\onlinecite{Paredes06} can again be mapped to a spin-one
chain in Eq.~(\ref{eqn:spinhamiltonian}) with $U=V=0$ and $n_0
=1$. That is, the lattice boson model studied in
Ref.~\onlinecite{Paredes06} is just a special case of the lattice
spin-one Hamiltonian studied in the present paper. Therefore, the
proposed gapless Pfaffian-like state should correspond to one of
the {\it gapless} ground states described by the present continuum
effective field theory. When $U=V=0$ and $n_0 =1$, ignoring the
$\xi$-dependent terms in Eq.~(\ref{eqn:spinhamiltonian}) as we did
before, the low-energy effective Hamiltonian becomes that in
Eq.~(\ref{Bosonform}) with the Luttinger parameters
$K_s=1$ and $K_a>1/2$. 
Thus it indicates that the gapless Pfaffian-like state should be
identified as the ground state laying on the SF-MI phase boundary,
where the charge sector is gapless and the $\Theta_a$ field is
pinned. As a nontrivial crosscheck of this observation, we show
that the correlation functions calculated by the Phaffian-like
wave function under a completely different approach can indeed be
reproduced within the present framework. By using the bosonized
forms Eq.~(\ref{boson-operator}) of the original lattice boson
operators, for the ground state laying on the SF-MI phase
boundary, the single-particle and boson-pair correlation functions
behave at long distance
\begin{equation}\label{Pfaffian-correlation}
\langle b_{i}^\dagger b_{j}\rangle\sim |i-j|^{-1/4K_s} , \quad
\langle (b_i^\dagger)^2 b_j^2 \rangle\sim |i-j|^{-K_s}
\end{equation}
with $K_s=1$. Notice that, on the SF-MI phase boundary, the
exponent of the single-particle correlation function is always
{\it four times larger} than that of the boson-pair correlation
function. These results agree with those in
Ref.~\onlinecite{Paredes06}. It supports our observation that the
ground state laying on the SF-MI phase boundary is nothing but the
Pfaffian-like state. Since a direct strong three-body interaction
among atoms are rare in nature,~\cite{Cooper} the authors in
Ref.~\onlinecite{Paredes06} suggested to use a Feschbach-resonant
system away from the resonance to generate an effective three-body
interaction among atoms, thus realizing this novel state.
According to the present discussion, we find that strong
three-body on-site repulsion is not essential for the appearance
of the Pfaffian-like state. Instead, 1D Interacting lattice boson
models can have a gapless Pfaffian-like ground state, as long as
they are tuned toward the critical points of the SF-MI phase
transition. Thus it is plausible to realize such 1D Pfaffian-like
states in an ordinary boson Hubbard model without borrowing the
mechanism of Feshbach resonance.

\section{Conclusions}

To summarize, a low-energy description, which successfully capture
the rich phases of the 1D lattice boson model near integer
fillings, is constructed. We believe that the present effective
theory provides general applications to related problems. For
example, parallel approaches can be used to study models of
coupled 1D hard-core lattice bosons, where similar effective
theories can be
constructed.~\cite{more_1d1,more_1d2,more_1d3,Mathey et
al,YWLee,unpublished} When local density fluctuations become
stronger, more local states with different particle numbers should
be kept. Following the above reasoning, an effective theory with
more bosonic phase fields should be reached. It is interesting to
see if more novel phases can be found in this case.

\begin{acknowledgments}
Y.-W. Lee, Y.-L. Lee, and M.-F. Yang acknowledge the support by
the National Science Council of Taiwan under the contracts NSC
96-2112-M-029-006-MY3, NSC 96-2112-M-018-006-MY2, and NSC
95-2112-M-029-006, respectively.

\end{acknowledgments}



\begin{thebibliography}{21}
\bibitem{comment1} As shown in Refs.~\onlinecite{Altman02} and
\onlinecite{Huber}, the derived theories under such a truncation
of the Hilbert space can even successfully describe the SF-MI
transition. Thus the restriction of allowing local particle number
fluctuation of order one should be sufficient to capture the
correct critical theory in the whole phase diagram.
\bibitem{asymmetry}
Their irrelevance can be realized as follows. First, the forms of
the leading nonlinear terms in our effective action
Eq.~(\ref{Bosonform}) are serverely constrained by the global U(1)
symmetry, the lattice translation symmetry and the
compactification radius of the bosonic phase fields. Therefore,
the $\xi$-dependent terms should not affect them. Second, while
bosonized terms like $\partial_x\Phi_s\partial_x\Theta_a$ or
$\partial_x\Phi_s$ are indeed generated by the $\xi$-dependent
terms, these bosonized terms will either modify the Luttinger
parameters (since the neutral mode is gapped) or their effects are
already included in the chemical potential term. Hence, we
conclude that the $\xi$-dependent terms will not make any
qualitative change for present discussions as long as $\xi\ll 1$.
\bibitem{bosonization}
We follow here the conventions used in Ref.~\onlinecite{Tsvelik},
where free fermions are mapped to bosonic phase fields with a
compactification radius $R=1/\sqrt{4\pi}$.
\bibitem{note1}
In fact, our Fig.~\ref{fig:fig1} is nothing but Fig.~1 in
Ref.~\onlinecite{Schulz86} after translating to the language of
lattice boson model. When more local states are kept, the phase
diagram will correspond to Fig.~2 in Ref.~\onlinecite{Schulz86}.
We notice that at a qualitative level, Fig.~2  and Fig.~1 of
Ref.~\onlinecite{Schulz86} are almost identical, i.e. the spin one
case and the other integer spin cases possess the same phases
apart from the slight changes of phase boundaries.
\bibitem{Ising} This can be most easily seen through a
``refermionization" procedure where we introduce a set of right-
and left-moving Dirac fermions $\chi_{R,L}=\frac{1}{\sqrt{2\pi
a}}e^{\pm i\sqrt{4\pi }\phi^\prime_{R, L}}$ with $\phi^\prime_{R,
L}=
\frac{1}{\sqrt{2K_a}}\Phi_a\mp\sqrt{\frac{K_a}{2}}\Theta_a=\Phi_a\mp\sqrt{\frac{1}{4}}\Theta_a$
to rewrite the Hamiltonian. For further details of the
bosonization description of the Ising transition, we refer the
reader to the book in Ref.~\onlinecite{Tsvelik}.
\bibitem{note2}
Since $\Phi_s$ is also pinned at
$\langle\Phi_s\rangle=\sqrt{\pi/8}$ in the CDW phase, the string
order parameter is also nonvanishing there, consistent with the
findings shown in Ref.~\onlinecite{Dalla Torre}.
\end{thebibliography}
\end{document}